\documentstyle[aps,preprint,tighten]{revtex}

\begin{document}

\draft

\title{Modern nucleon-nucleon interactions and charge-symmetry breaking in 
nuclei}

\author{C.\ Harzer, H.\ M\"{u}ther}
\address{Institut f\"ur Theoretische Physik, Universit\"at T\"ubingen,
         D-72076 T\"ubingen, Germany}
\author{R.\ Machleidt}
\address{Department of Physics, University of Idaho, Moscow,
         Idaho 83844, U.S.A.}

\maketitle

\begin{abstract}
Coulomb displacement energies, i.e., the differences between the energies of
corresponding nuclear states in mirror nuclei,  are evaluated using recent
models for the nucleon-nucleon (NN) interaction.  These modern NN potentials
account for breaking of isospin symmetry and reproduce $pp$ and $pn$ phase
shifts accurately.  The predictions by these new potentials for the binding of
$^{16}O$ are calculated. A particular focus of our study are effects due to
nuclear correlations and charge-symmetry breaking (CSB).  We find that the CSB
terms in the modern NN interactions substantially reduce the discrepancy
between theory and experiment for the Coulomb displacement energies; however,
our calculations do not completely explain  the Nolen-Schiffer anomaly. 
Potential sources for the remaining discrepancies are discussed.
\end{abstract}


The differences between the energies of corresponding states in mirror nuclei,
the so-called Coulomb displacement energies, are due to
charge-symmetry breaking (CSB) of the nucleon-nucleon (NN) interaction. If one
assumes that the strong part of the nuclear force is charge symmetric, 
i.e. the strong proton-proton interaction is identical to the interaction  
between two neutrons, then the Coulomb
displacement energies would originate entirely from the electromagnetic
interaction between the nucleons. The dominant contribution is the Coulomb
repulsion. After accurate experimental data on the charge distribution became
available from electron scattering experiments, Hartree-Fock calculations with
phenomenological models for the NN interaction like the Skyrme forces were
performed which reproduced these measured charge distributions with good
accuracy. The Coulomb displacement energies which were evaluated with 
these Hartree-Fock wave functions, however, underestimated the experimental data
by typically seven percent. This has become known as the Nolen-Schiffer
anomaly\cite{nolen}. Many attempts have been made to explain this discrepancy 
by the inclusion of electromagnetic corrections, many-body correlations beyond
the Hartree-Fock approach, or by explicit charge-symmetry breaking terms in the
NN interaction\cite{miller,auerb,sato,blunden,kuo}.

During the last few years, a new generation of realistic NN interactions has
been developed, which yield very accurate fits of the proton-proton (pp) and
proton-neutron (pn) data\cite{nim,v18,cdbonn}. These new
interactions account for isospin symmetry breaking (ISB) 
and also for CSB (which is a special case of ISB).  The
long-range part of these interactions is described in terms of the
one-pion-exchange model, accounting correctly
for the mass difference between the neutral pion, $\pi^0$, and
the charged pions, $\pi^\pm$. This distinction between the masses of
a neutral and charged pions is one origin for ISB in the resulting NN
interactions. Moreover, these interactions also account 
for the mass difference between
proton and neutron. This gives rise to a difference in the matrix elements of
the meson-exchange diagrams between two protons as compared to two neutrons. 
Within the one-boson-exchange model, this yields only a very small
contribution that breaks charge symmetry.

Besides the latter effect,
the Argonne and Bonn potentials (which we will apply here) 
include additional CSB terms necessary 
to correctly reproduce the empirical differences in the scattering length and 
effective range parameters for $pp$ and $nn$ scattering in the $^1S_0$
state. The Argonne $V_{18}$
(AV18) potential\cite{v18} is constructed in a ($S,T$) decomposition (where 
$S$ and $T$ denote the total spin and isospin of the two interacting 
nucleons). The local
potentials in the ($S=0,T=1$) channel are adjusted such as to 
reproduce the $^1S_0$ scattering
length for the various isospin projections. This method of constructing
CSB potentials implies that the information on
CSB from the $^1S_0$ scattering length and effective range parameters 
is simply extrapolated
to channels with $L>0$. 

More reliably, the information on CSB in NN partial waves with $L>0$
can be derived from a comprehensive meson-exchange model that includes
diagrams beyond the simple one-meson-exchange approximation.
Based upon the Bonn Full Model~\cite{elster},
ISB effects due to hadron mass-splitting have carefully been calculated 
up to partial waves with $J=4$
in ref.~\cite{LM98}.
The new high-precision NN potential CDBonn99~\cite{cdbo99}
includes these ISB effects plus the effects from irreducible 
$\pi\gamma$ exchange as derived in \cite{kolk}. 
The difference between CDBonn99 and 
CDBonn96~\cite{cdbonn}
is that the latter takes CSB only in $^1S_0$ into account and
not in higher partial waves.
Thus, a comparison between CDBonn99 and CDBonn96 demonstrates the
effect from CSB in states with $L>0$.
This will be useful for our discussion below.

It is the aim of the present work to investigate the predictions by these new
potentials for the properties of finite nuclei. Our example nucleus is
$^{16}O$. In particular, we want to determine the effects of correlations and
of CSB by these potentials on the calculated Coulomb displacement energy.
One possibility would be to perform self-consistent
Brueckner-Hartree-Fock (BHF) calculations and extract the Coulomb displacement
energies from the single-particle energies for protons and neutrons. We do not
take this approach, for the following reasons: (i) Such self-consistent BHF
calculations typically predict the radii for the charge-density distributions
too small\cite{carlo}. This implies
that the leading Coulomb contribution to the displacement energy would be
overestimated. Also the calculation of the correction terms would be based on
single-particle wavefunctions which are localized too much. (ii) BHF
calculations are appropriate for short-range correlations. However,
long-range correlations involving the admixture of configurations with low
excitation energies in the uncorrelated shell-model basis require a more
careful treatment. (iii) The BHF single-particle energies do not account for
any distribution of the single-particle strength consistent with
realistic spectral functions.

For the reasons listed,
we take the following approach. We use single-particle wave functions
from Hartree-Fock calculations with effective nuclear forces, which yield
a good fit to the empirical charge distribution. These wave functions
are used to determine the leading Coulomb contribution and corrections like the
effects of finite proton size, the electromagnetic spin-orbit interaction, the
kinetic energy correction due to the mass difference between proton and neutron,
and the effects of vacuum polarization. Actually, for these contributions we
use the results by Sato\cite{sato}. The first column of our table 2 is 
taken from table 2 of ref.\cite{sato} which includes 
all the effects just listed.

The correlation effects are taken into account in a two-step procedure. 
We assume a
model space defined in terms of shell-model configurations
including oscillator single-particle states up to the 1p0f shell. 
We use the oscillator parameter $b=1.76$ fm which is appropriate for
$^{16}O$. The effects of short-range correlations are calculated
by employing an effective interaction,  i.e.\ a ${\cal G}$-matrix 
suitable for the model
space. This ${\cal G}$-matrix is determined as the solution of the
Bethe-Goldstone equation
\begin{equation}
{\cal G}(\Omega ) =  V + V \frac{Q_{\hbox{mod}}}{ \Omega - Q_{\hbox{mod}} T
Q_{\hbox{mod}}} {\cal G} (\Omega )\; ,
\label{gmat}
\end{equation}
where $T$ is identified with the kinetic energy operator, while $V$
stands for the bare two-body interaction including the Coulomb interaction and
accounting for ISB terms in the strong interaction. The Pauli operator
$Q_{\hbox{mod}}$  in this Bethe-Goldstone eq.(\ref{gmat}) is defined in terms
of two-particle harmonic oscillator states $\vert \alpha\beta >$ by
\begin{equation}
 Q_{\hbox{mod}} \vert \alpha\beta > = \left\{ \begin{array}{ll}
0 & \mbox{if $\alpha$ or $\beta$ from $0s$ or $0p$ shell}\cr
0 & \mbox{if $\alpha$ and $\beta$ from $1s0d$ or $1p0f$ shell} \cr
\vert \alpha\beta > & \mbox{elsewhere} \end{array} \right.
\label{paul}
\end{equation} 
As a first approximation we use the resulting ${\cal G}$-matrix elements and
evaluate single-particle energies in the BHF approximation
$\epsilon_{\alpha}$. This
approximation, which will be denoted as BHF in the discussion below, 
accounts for short-range correlations, which are described in terms of
configurations outside our model space. In a next step we add to this BHF
definition of the nucleon self-energy the irreducible terms of second order in
${\cal G}$ which account for intermediate two-particle one-hole and one-particle
two-hole configurations within the model-space
\begin{equation}
{\cal U}_{\alpha}^{(2)} = \frac{1}{2} \sum_{p_1,p_2,h}\frac{<\alpha h \vert
 {\cal G}\vert p_1p_2><p_1p_2\vert {\cal G}\vert \alpha h>}{\omega -
(\epsilon_{p_1} + \epsilon_{p_2}-\epsilon_h) + i\eta} +
\frac{1}{2} \sum_{h_1,h_2,p}\frac{<\alpha p \vert
 {\cal G}\vert h_1h_2><h_1h_2\vert {\cal G}\vert \alpha h>}{\omega -
(\epsilon_{h_1} + \epsilon_{h_2}-\epsilon_p) - i\eta}\,.
\end{equation}
 Applying the techniques
described in \cite{skour} we can solve the Dyson equation for the
single-particle Greens function $G_{\alpha\beta}(\omega )$
\begin{equation}
G_{\alpha}(\omega) = g^{\alpha}(\omega) + 
g_{\alpha}(\omega){\cal U}_{\alpha}^{(2)}(\omega) G_{\alpha}(\omega)
\end{equation}
with $g_\alpha$ the BHF approximation for the single-particle Greens function, 
and determine its Lehmann representation
\begin{equation}  
G_{\alpha}(\omega ) = \sum_n \frac{\left| <\Psi_n^{A+1} \vert 
a^\dagger_\alpha \vert \Psi_0 >\right|^2} {\omega - (E_n^{A+1}-E_0) + i\eta}
+  \sum_m \frac{\left| <\Psi_m^{A-1} \vert 
a_\alpha \vert \Psi_0 >\right|^2} {\omega - (E_0 -E_m^{A-1}) - i\eta}\,.
\label{lehm}
\end{equation}
This yields directly the energies of the states with $A\pm 1$ nucleons we are
interested in, as well as the spectroscopic factors for nucleon addition or
removal. 

Numerical results for the binding energy of $^{16}O$ and some single-particle
properties are listed in table 1. This table contains two columns for each NN
interaction considered. The columns labeled ``ISB'' contain results of
calculations in which the ISB terms of the interactions are taken into
account. In order to demonstrate the size of these ISB terms we also performed
calculations, in which we restored isospin-symmetry of the strong interaction by
replacing the $pp$ and $nn$ interactions by the corresponding $pn$ interaction
(see columns labeled $pn$).

It is worth noting that the inclusion of long range correlations by means of the
Greens function approach outlined above yields an additional binding energy of
around 2 MeV per nucleon for all interactions considered. About 1.5 MeV of
these 2 MeV per nucleon can be attributed to the admixture of the low-lying
particle-particle states within the model space. This energy would also be
included in a BHF calculation using the BHF Pauli operator in the
Bethe-Goldstone eq.(\ref{gmat}) instead of $Q_{\hbox{mod}}$ defined in
(\ref{paul}). An additional 0.5 MeV per nucleon arises from the inclusion of the
hole-hole scattering terms in the Greens function approach. 

The effects of long range correlations are also very important for the
quasiparticle energies. For this quantity, however, the effects of low energy
particle-particle and hole-hole contributions tend to cancel each other to a
large extent. The inclusion of 2p1h configurations within the model space lowers
the proton quasiparticle energy for the $p_{1/2}$ state (using AV18) from
-12.12 MeV to -15.38 MeV. If, however, the admixture of 2h1p configurations is
also included in the definition of the self-energy, the quasiparticle 
energy yields -12.54 MeV, close to its original BHF value. Similar
cancellations are observed for other states and, also, using other 
NN interactions.

The Bonn potentials CDBonn96 and CDBonn99 yield around 0.9 MeV per nucleon more
binding energy than the Argonne potential AV18. This is to be compared with a
difference of 1.2 MeV per nucleon, which has been obtained comparing the results
of BHF calculations for these potentials in nuclear matter at saturation
density\cite{pols}. These energy differences can be related to the fact that the
Argonne potential is ``stiffer'' than the 
Bonn potential\cite{pols,mort98}. More stiffness creates more
correlations.  This can be seen from the spectroscopic
factors in table 1, which deviate more from unity
in the case of AV18 as compared to Bonn. 

Including the ISB terms in the NN interaction rather than using the $np$
interaction for all isospin channels has a small but non-negligible effect on
the calculated binding energies. The scattering lengths
for $pp$ and $pn$ scattering implies an NN interaction which is slightly more
attractive in the $pn$ as compared to $pp$. This small difference
translates into about 0.2 MeV per nucleon in $^{16}O$.

We finally discuss the effects of correlations and of charge-symmetry
breaking in the NN interaction on the calculated Coulomb displacement energies.
Results are listed in table 2 for various one-hole and one-particle states
relative to $^{16}O$. The first column of this table, $C^{(1)}$,
contains the results of
ref.\cite{sato} for the leading Coulomb contributions, 
the corrections due to the
finite proton size, the electromagnetic spin-orbit interaction, the
kinetic energy correction due to nucleon mass splitting,
and the effects of vacuum polarization. As discussed above, we think
that it is more realistic to evaluate these contributions for single-particle
wave functions which are derived from Hartree-Fock calculations with
phenomenological forces rather than using the wavefunctions derived from a
microscopic BHF calculation.

The second and third columns of table 2 list the corrections to the Coulomb
displacement energies which originate from the treatment of short-range
($\delta_{SR}$) and long-range correlations ($\delta_{LR}$) discussed above. The
correction $\delta_{SR}$ has been derived from the differences of
BHF single-particle energies for protons and neutrons subtracting the Coulomb
displacement energy evaluated in the mean-field approximation
\begin{equation}
\delta_{SR} = \epsilon_i^{BHF} (\mbox{proton}) - \epsilon_i^{BHF}
(\mbox{neutron}) - \delta_{\mbox{mean field}} 
\end{equation}
In this case the BHF calculations have been performed with the $pn$ versions of
the different interactions, i.e.~without any ISB terms. The correction terms
$\delta_{LR}$ have been evaluated in a similar way from the quasiparticle
energies determined in the Greens function approach, subtracting the BHF effects
already contained in $\delta_{SR}$. The correction terms $\delta_{SR}$ and
$\delta_{LR}$ include the effects represented by irreducible diagrams of second
and higher order in the interaction, in which at least one of the interaction
lines represents the Coulomb interaction. In addition they contain the effects 
of folded diagrams discussed by Tam et al.\cite{kuo}. We find that the
correlation effects are rather weak. The short- and long-range contributions
tend to cancel each other. This is true in particular for the one-hole states
$p_{3/2}^{-1}$ and $p_{1/2}^{-1}$. The effects of short-range correlations
dominate in the case of the particle states, $d_{5/2}$ and $1s_{1/2}$, leading
to a total correlation effect around 100 keV in the Coulomb displacement
energies. This effect is slightly larger for the Argonne potential than for the
Bonn potentials because of the stronger correlations in the case of Argonne.

The contributions to the Coulomb displacement energies caused by the CSB
terms in the NN interactions, $\delta_{CSB}$, 
are listed in the fourth column of table 2.
The CDBonn96 potential, which includes CSB only in the $^1S_0$ state
to fit the empirical $pp$ and $nn$ scattering lengths,
provides the smallest $\delta_{CSB}$.
When CSB as derived from a comprehensive meson-exchange model
is included in partial waves with $L>0$, as done in CDBonn99, then
the $\delta_{CSB}$ contribution about doubles. The Argonne AV18
potential also includes CSB for $L>0$ and, therefore, produces
a relatively large $\delta_{CSB}$.
Note, however, that the CSB for $L>0$ in AV18 is just an extrapolation
of what $^1S_0$ needs to fit the $pp$ and $nn$ scattering lengths;
it is not based upon theory.

In any case, when a NN potential includes CSB beyond the 
$^1S_0$ state, then a contribution to the Coulomb
displacement energies of about 100 keV is created, while
CSB in $^1S_0$ only generates merely about 50 keV.
This demonstrates how important it is to include CSB in all
relevant NN partial waves if one wants to discuss a phenomenon
like the Nolen-Schiffer anomaly in a proper way.

However, it also turns out that even this carefull consideration
of CSB does not fully explain the Nolen-Schiffer anomaly, in our
calculations.
Our final predictions given in column $C^{Tot}$ of table 2 still
differ by about 100 keV from the experimental values.
Thus the CSB NN force contribution has cut in half the original discrepancy of
about 200 keV.

For the remaining discrepancy, many explanations are possible.
First, the nuclear structure part of our calculations may carry
some uncertainty. To obtain an idea of how large such uncertainties
may be, we compare the results for Coulomb displacement energies
using the Skyrme II force and {\it no CSB} by Sato~\cite{sato}
with the more recent ones by Suzuki {\it et al.}~\cite{SSA92}.
For the single-hole state $p^{-1}_{1/2}$, 
Suzuki's result is larger by 167 keV as compared to Sato; 
and for the single-particle state $d_{5/2}$, the two calculations
differ by 138 keV.
Uncertainties of this size can well explain the remaining discrepancies
in our results.

Another possibility is that the CSB forces contained in CDBonn99
and AV18 are too weak. Based upon our results, it may be 
suggestive to conclude that
CSB forces of about twice their current strength are needed.
Notice, however, that one cannot just add more CSB forces
to these potentials. A crucial constraint
for any realistic CSB NN force is that it reproduces the empirical
difference between the $pp$ and $nn$ $^1S_0$ scattering lengths,
$\Delta a_{CSB} 
= 1.5 \pm 0.5$ fm~\cite{miller}.
The CSB contained in CDBonn99 is based upon nucleon mass-difference effects
as obtained in a comprehensive meson-exchange model which completely
explains the entire $\Delta a_{CSB}$~\cite{LM98} leaving no room 
for additional CSB contributions.

The only possibility that remains then 
is to simply ignore the above
CSB effects and consider an alternative source for CSB, namely
$\rho^0-\omega$ mixing.
Traditionally, it was believed that 
$\rho^0-\omega$ 
mixing causes essentially all CSB in the nuclear force~\cite{miller}.
However, recently some doubt has been cast on this paradigm.
Some researchers~\cite{GHT92,PW93,KTW93} found that 
$\rho^0-\omega$ exchange may have a substantial
$q^2$ dependence such as to cause this contribution to nearly vanish
in $NN$.
The recent findings of ref.~\cite{LM98} that the empirically known CSB in the 
nuclear force can be explained solely from nucleon mass splitting 
(leaving essentially no room for additional CSB contributions 
from $\rho^0-\omega$ mixing or other sources) fits well into this 
new scenario.
However, since the issue of the $q^2$ dependence of
$\rho^0-\omega$ exchange and its impact on $NN$ is by no means settled
(see Refs.~\cite{MO95,CMR97} for critical discussions and more references), 
it is premature to draw any definite conclusions.
In any case, for test purposes one may invoke the 
$\rho^0-\omega$ 
mechanism as an alternative.

Note, however, that due to the constraint that the $^1S_0$
$\Delta a_{CSB}$ be reproduced quantitatively, 
the $^1S_0$ contribution will most likely not change,
no matter what miscroscopic mechanism is assumed for CSB.
However, the CSB contributions in partial waves with
$L>0$ may depend sensitively on the underlying mechanism.
The CSB force caused by nucleon mass-splitting has essential
scalar character, while
$\rho^0-\omega$ 
exchange is of vector nature.
Since we have seen above that the $L>0$ partial waves
produce about 50\% of the total CSB effect,
higher partial waves may carry the potential for substantial changes.
This would be an interesting topic for a future investigation.

In summary,
we have compared results for bulk properties of finite nuclei derived
from modern models of the nucleon-nucleon interaction. Effects of short-range as
well as long-range correlations are taken into account. The different models
for the NN interaction are essentially phase-shift equivalent. Nevertheless 
they predict differences in the binding energy of $^{16}O$ up to 1 MeV per
nucleon. The main source for this discrepancy could be the local versus 
non-local description of the pion exchange interaction as discussed
in the literature for the deuteron\cite{deuter} 
and nuclear matter\cite{mort98}.
The CSB force components contained in the CDBonn99 and AV18 potentials
cut in half the discrepancy that is know as the Nolen-Schiffer
anomaly. The remainder of the discrepancy may be due to subtle
nuclear structure effects left out in our current calculations.
The consideration of alternative mechanism for CSB has also the
potential to shed light on open issues.

This work was supported in part by the Graduiertenkolleg ``Struktur und 
Wechselwirkung von Hadronen und Kernen'' (DFG, GRK 132/3) and by the
U.S.~National Science Foundation
under Grant No.\ PHY-9603097.

\begin{table}
\begin{tabular}{c|rrrrrr}
& \multicolumn{2}{c}{CDBonn96} & \multicolumn{2}{c}{CDBonn99} & \multicolumn{2}{c}
{AV18}\\
& $pn$ & ISB & $pn$ & ISB & $pn$ & ISB \\
&&&&&&\\
\hline
&&&&&&\\
$\epsilon^{BHF}_{p3/2}$ [MeV] & -17.37 & -17.23 & -17.39 & -17.05 & -15.89 &
-15.58 \\
$\epsilon^{BHF}_{p1/2}$ [MeV] & -13.68 & -13.49 & -13.71 & -13.36 & -12.43 &
-12.12 \\
$\epsilon^{BHF}_{d5/2}$ [MeV] &   0.39 &   0.99 &   0.36 & 0.99 & 1.34 & 2.07 \\
$E^{BHF}/A$ [MeV] & -4.80 & -4.72 & -4.81 & -4.65 & -3.97 & -3.82 \\
&&&&&&\\
$\epsilon^{QP}_{p3/2}$ [MeV] &  -17.65 & -17.52 & -17.68 & -17.32 & -16.14 &
-15.82 \\
$\epsilon^{QP}_{p1/2}$ [MeV] &  -14.23 & -14.03 & -14.26 & -13.88 & -12.88 &
-12.54 \\
$\epsilon^{QP}_{d5/2}$ [MeV] &   -0.57 &   0.05 &  -0.60 & 0.03 & 0.44 & 1.16 \\
&&&&&&\\
$S_{p3/2}$ & .8071 & .8089 & .8070 & .8087 & .7867 & .7852 \\
$S_{p1/2}$ & .7937 & .7960 & .7936 & .7956 & .7721 & .7745 \\
$S_{d5/2}$ & .8460 & .8474 & .8460 & .8472 & .8238 & .8261 \\
&&&&&&\\
E/A [MeV] & -6.89 & -6.70 & -6.90 & -6.70 & -5.94 & -5.76 \\
\end{tabular}

\caption{Single-particle properties and binding energy per nucleon of $^{16}O$
calculated for the potentials CDBonn96, CDBonn99, and AV18. The results in columns 
ISB are obtained by taking the ISB of these potentials properly into
account while for columns $pn$ only the $pn$
potentials are used throughout ignoring ISB.
Results for proton single-particle energies 
are listed for the BHF approximation 
($\epsilon_\alpha^{BHF}$)
and for the complete Greens
function approach 
($\epsilon_\alpha^{QP}$).
For the latter case,
also the spectroscopic factors $S_\alpha$
are listed. The total energies per nucleon (denoted by $E^{BHF}/A$ for BHF and
by $E/A$ for the Greens function approach) 
include a correction for the spurious center of mass motion.}
\end{table}

\begin{table}
\begin{tabular}{cc|rrrrrr}
&& $C^{(1)}$ & $\delta_{SR}$ & $\delta_{LR}$ & $\delta_{CSB}$ & $C^{Tot}$ & 
Exp \\
\hline
$p_{3/2}^{-1}$ & CDBonn96 & 3205 &  -44 & 46 & 36 & 3240 & 3395 \\
& CDBonn99 & & -44 & 46 & 86 & 3290 & \\
& AV18 & & -71 & 47 & 108 & 3285 & \\
\hline
$p_{1/2}^{-1}$ & CDBonn96 & 3235 & -52 & 37 & 54 & 3271 & 3542 \\
& CDBonn99 & & -52 & 37 & 91 & 3320 & \\
& AV18 & & -79 & 39 & 103 & 3297 & \\
\hline
$d_{5/2}$ & CDBonn96 & 3135 & 154 & -15 & 49 & 3326 & 3542 \\
& CDBonn99 & & 154 & -15 & 72 & 3350 & \\
& AV18 & & 187 & -18 & 92 & 3401 & \\
\hline 
$1s_{1/2}$ & CDBonn96 & 2905 & 159 & -45 & 58 & 3081 & 3166 \\  
& CDBonn99 & & 160 & -46 & 93 & 3117 & \\
& AV18 & & 198 & -47 & 112 & 3174 & \\
\end{tabular}

\caption{ Coulomb displacement energies for single-hole ($p_{3/2}^{-1}$ and 
$p_{1/2}^{-1}$) and single-particle states ($d_{5/2}$ and $1s_{1/2}$) around
$^{16}O$. The single-particle contribution, $C^{(1)}$, is from Sato
\protect\cite{sato}. Contributions due to short-range correlations,
$\delta_{SR}$, long-range correlations inside the model space, $\delta_{LR}$,
and due to the charge-symmetry breaking terms in the strong interaction,
$\delta_{CSB}$, are
calculated for three different NN interactions. 
The total results for the displacement energies, $C^{Tot}$, are compared to
the experimental data given in the last column. All entries are in keV.}
\end{table}
\end{document}